# Creation and unification of development and life stage ontologies for animals

Anne Niknejad[1,2], Christopher J. Mungall[3], David Osumi-Sutherland[4], Marc Robinson-Rechavi[1,2], Frederic B. Bastian[1,2].

[1]Department of Ecology and Evolution, University of Lausanne, Switzerland
[2]SIB Swiss Institute of Bioinformatics, Lausanne, Switzerland
[3]Lawrence Berkeley National Laboratory, Berkeley, USA
[4]EMBL-European Bioinformatics Institute, Wellcome Genome Campus, Hinxton, Cambridge, UK

**ABSTRACT**
**Motivation:** With the new era of genomics, an increasing number of animal species are amenable to large-scale data generation. This had led to the emergence of new multi-species ontologies to annotate and organize these data. While anatomy and cell types are well covered by these efforts, information regarding development and life stages is also critical in the annotation of animal data. Its lack can hamper our ability to answer comparative biology questions and to interpret functional results. We present here a collection of development and life stage ontologies for 21 animal species, and their merge into a common multi-species ontology. This work has allowed the integration and comparison of transcriptomics data in 52 animal species.

## 1 INTRODUCTION

In the past decades, the field of functional biology has moved from the study of a core set of model organisms, in the direction of the end goal of studying all species, notably thanks to the advances in genetic sequencing technologies. This drove the progress of multi-species ontology development and usage, to capture structured information in a multi-species manner. For instance, the Uberon ontology allows to describe the anatomy of any animal species (Haendel et al., 2014), and the project PANTHER aims at classifying protein-coding genes in all domains of life, notably by applying Gene Ontology annotations on protein evolutionary trees (Mi et al., 2021).

Applying the same principles to describe development and life stages in animals is a challenging task. First, different species might not undergo the same developmental stages, such as a pupal stage existing only in Holometabola. Second, organs do not develop at the same speed and with the same sequence in different species, i.e. development is heterochronous (e.g., (Jeffery et al., 2005)). Third, even for very similar stages, such as "gastrula" in Eumetazoa, the duration varies between species. As a result, ontologies have been developed to describe the development of one species or of a limited set of species at a time, such as: EMAPA for mouse (Hayamizu et al., 2015); FBdv for fly (Costa et al., 2013); ZFA for zebrafish (Van Slyke et al., 2014); XAO for Xenopus (Segerdell et al., 2008, 2013); and WBls (Harris et al., 2019) for *C. elegans*.

We present here how we have overcome these challenges to scale up the annotations of phenotypic and genomic data to an increasing number of animal species. We have adopted a dual approach of describing the development and life stages of individual species, while using as backbone a core multi-species structure, allowing to align them all. This approach has notably been used to allow the integration of transcriptomics data for 52 animal species in the gene expression database Bgee (Bastian et al., 2021). All the data are available at https://github.com/obophenotype/developmental-stage-ontologies, and the multi-species unification is available directly from Uberon (see, e.g., http://purl.obolibrary.org/obo/uberon/subsets/life-stages-composite.obo).

## 2 METHODS

Our approach has been to develop single-species ontologies, to capture with fine-grained details the development and life stages of individual species. These single-species ontologies are then merged into a multi-species composite stage ontology. Having one common multi-species ontology, with high level terms merged between species (e.g. 'gastrula'), allows us to integrate data to comparable stages in different species, and thus take development into account in comparative functional genomics.

### 2.1 Multi-species core development and life stage ontology

The core multi-species ontology has been developed by the Uberon maintainers. It includes high level terms, generic to several animal species, with 'part_of', 'preceded_by', and 'immediately_preceded_by' object properties between classes. Taxon constraints (Deegan (née Clark) et al., 2010; Kuśnierczyk, 2008) are used to restrict stage existence in some taxa, for instance, to indicate that a neurula stage only exists in Eumetazoa. This core ontology is available at http://purl.obolibrary.org/obo/uberon/subsets/life-stages-minimal.obo.

### 2.2 Species-specific ontologies

For 21 of the 52 species integrated in Bgee, we have developed precise species-specific ontologies describing the development and life stages of each of them (see Table 1 for details). This includes notably precise data properties in days, months or years to capture the duration of each stage. This is a difficult task, as we try here to capture the continuous process of development as a set of discrete stages. This is exemplified by the problem of the age of weaning in humans: we have defined the stage HsapDv:0000260 "nursing stage" as stopping at 1 year-old, while there exist large variations due to cultural and individual differences. The "natural" age of weaning in humans could be of several years according to some authors (see, e.g., (Leeson, 2001). In such cases, we carefully review the literature and choose the best compromise, and provide the sources for our choices as comments in the ontology.

For fly, zebrafish, Xenopus, and C. elegans, we rely on the ontologies developed by Model Organism Databases (see Introduction). For the remaining 27 species integrated in Bgee, we rely on the core multi-species ontology developed by Uberon. For these species, we don't have data available requiring fine-grained annotations of development and life stages, so that the high level terms are sufficient and allow data integration.

### 2.3 Merge of newly created and already existing ontologies

For each species-specific ontology, we create a "bridge" file to the Uberon core multi-species structure. This file provides relations for some of the species-specific terms to the Uberon multi-species terms which are used in their logical definition, and also allows to add relationships to attach them to the Uberon graph. These bridge files exist for the 21 species integrated with fine-grained details, and the 5 model organisms (see Introduction). For instance, the human bridge file specifies that the class HsapDv:0000037 "fetal stage" is equivalent to the intersection of the named class UBERON:0007220 "late embryonic stage", and of the anonymous class "(part_of some NCBITaxon:9606 Homo sapiens)".

Thanks to these bridge files, all the species-specific ontologies can be automatically merged to generate a composite multi-species ontologies, merging in taxonomic equivalents, and relabeling species-specific classes. This allows to use both the high level terms common to multiple species, such as UBERON:0000109 "gastrula stage", and highly precise species-specific terms, such as HsapDv:0000011 "Carnegie stage 06 (human)", which becomes after the merge "part_of" UBERON:0000109 "gastrula stage".

## 3 RESULTS

### 3.1 Ontology statistics

Table 1 presents the 21 species with species-specific ontologies created for this project. The differences in the number of classes linked to Uberon is related to the species-specific ontology granularity, the known biology of the species, and how well the ontology aligns with Uberon. These species-specific ontologies are available at https://github.com/obophenotype/developmental-stage-ontologies/tree/master/src . A human-friendly report of the result of the merge for the 52 species integrated in Bgee is available at https://github.com/obophenotype/developmental-stage-ontologies/blob/master/external/bgee/report.md. The resulting merged composite ontology is available at http://purl.obolibrary.org/obo/uberon/subsets/life-stages-composite.obo.

**Table 1.** Details of the 21 species with species-specific ontologies created for this project

| Ontology name | Species | Number of classes | Number of classes linked to Uberon |
|---|---|---|---|
| AcarDv | Anolis carolinensis | 20 | 20 |
| BtauDv | Bos taurus | 78 | 8 |
| CfamDv | Canis lupus familiaris | 24 | 24 |
| CporDv | Cavia porcellus | 24 | 24 |
| DpseDv | Drosophila pseudoobscura | 21 | 18 |
| DsimDv | Drosophila simulans | 21 | 18 |
| EcabDv | Equus caballus | 24 | 24 |
| FcatDv | Felis catus | 24 | 24 |
| GgalDv | Gallus gallus | 90 | 16 |
| GgorDv | Gorilla gorilla | 47 | 24 |
| HsapDv | Homo sapiens | 238 | 20 |
| MdomDv | Monodelphis domestica | 59 | 24 |
| MmulDv | Macaca mulatta | 66 | 24 |
| MmusDv | Mus musculus | 134 | 20 |
| OanaDv | Ornithorhynchus anatinus | 24 | 24 |
| OariDv | Ovis aries | 33 | 20 |
| OcunDv | Oryctolagus cuniculus | 24 | 24 |
| PpanDv | Pan paniscus | 56 | 24 |
| PtroDv | Pan troglodytes | 67 | 24 |
| RnorDv | Rattus norvegicus | 60 | 20 |
| SscrDv | Sus scrofa | 94 | 22 |

Namespace of each ontology, species name of the related species, total number of classes in the ontology, and total number of classes linked to the multi-species core Uberon ontology.

### 3.2 Application examples

As mentioned earlier, the resulting merged multi-species ontology has allowed the integration of 52 animal species into the gene expression database Bgee. Bgee annotates each transcriptomics sample (e.g., RNA-Seq library) to anatomy, development and life stage, sex, and strain. The work presented here allows to capture metadata information regarding development and life stage in a structured way. It also allows comparison of expression patterns between species, taking into account development. While it is not possible to compare data between precise species-specific stages (see Introduction), propagating these data along the graph of the merged multi-species ontology allows comparison at a broad level, such as "gastrula" or "blastula".

Another application is the use of species-specific ontologies in projects dedicated to studying one species. This is for instance the case in the Human Cell Atlas project (Rozenblatt-Rosen et al., 2017), where development and life stage information is captured by using the HsapDv ontology described in this paper.

## 4 CONCLUSIONS

Because development varies greatly between species, notably the duration and sequences of stages, we have chosen the pragmatic approach of capturing precise species-specific information when it is necessary to capture available metadata. However, for datasets from most species, the information regarding development and life stages is not precise enough to be worth developing a new ontology; in that case, the core multi-species structure of Uberon is sufficient. Adopting this dual approach allowed us to scale up the integration of available data in many animal species. We intend to continue developing new species-specific ontologies as required by available data, and merging them in the core structure of Uberon, allowing comparative biology analyses.

## ACKNOWLEDGEMENTS

We thank Jason Hilton and Ilinca Tudose for comments and bug reports on the developmental stage ontologies tracker. Amos Bairoch's comments were critical for updates to the human ontology HsapDv. We thank all the Model Organism Databases that developed the core developmental stage ontologies used for model organisms.

## REFERENCES

Bastian, F. B., Roux, J., Niknejad, A., Comte, A., Fonseca Costa, S. S., de Farias, T. M., Moretti, S., Parmentier, G., de Laval, V. R., Rosikiewicz, M., Wollbrett, J., Echchiki, A., Escoriza, A., Gharib, W. H., Gonzales-Porta, M., Jarosz, Y., Laurenczy, B., Moret, P., Person, E., … Robinson-Rechavi, M. (2021). The Bgee suite: Integrated curated expression atlas and comparative transcriptomics in animals. Nucleic Acids Research, 49(D1), D831–D847.

Costa, M., Reeve, S., Grumbling, G., & Osumi-Sutherland, D. (2013). The Drosophila anatomy ontology. Journal of Biomedical Semantics, 4(1), 32.

Deegan (née Clark), J. I., Dimmer, E. C., & Mungall, C. J. (2010). Formalization of taxon-based constraints to detect inconsistencies in annotation and ontology development. BMC Bioinformatics, 11(1), 530.

Haendel, M. A., Balhoff, J. P., Bastian, F. B., Blackburn, D. C., Blake, J. A., Bradford, Y., Comte, A., Dahdul, W. M., Dececchi, T. A., Druzinsky, R. E., Hayamizu, T. F., Ibrahim, N., Lewis, S. E., Mabee, P. M., Niknejad, A., Robinson-Rechavi, M., Sereno, P. C., & Mungall, C. J. (2014). Unification of multi-species vertebrate anatomy ontologies for comparative biology in Uberon. Journal of Biomedical Semantics, 5(1), 21.

Harris, T. W., Arnaboldi, V., Cain, S., Chan, J., Chen, W. J., Cho, J., Davis, P., Gao, S., Grove, C. A., Kishore, R., Lee, R. Y. N., Muller, H.-M., Nakamura, C., Nuin, P., Paulini, M., Raciti, D., Rodgers, F. H., Russell, M., Schindelman, G., … Sternberg, P. W. (2019). WormBase: A modern Model Organism Information Resource. Nucleic Acids Research, gkz920.

Hayamizu, T. F., Baldock, R. A., & Ringwald, M. (2015). Mouse anatomy ontologies: Enhancements and tools for exploring and integrating biomedical data. Mammalian Genome, 26(9–10), 422–430.

Jeffery, J. E., Bininda-Emonds, O. R. P., Coates, M. I., & Richardson, M. K. (2005). A New Technique for Identifying Sequence Heterochrony. Systematic Biology, 54(2), 230–240.

Kuśnierczyk, W. (2008). Taxonomy-based partitioning of the Gene Ontology. Journal of Biomedical Informatics, 41(2), 282–292.

Leeson, C. P. M. (2001). Duration of breast feeding and arterial distensibility in early adult life: Population based study. BMJ, 322(7287), 643–647.

Mi, H., Ebert, D., Muruganujan, A., Mills, C., Albou, L.-P., Mushayamaha, T., & Thomas, P. D. (2021). PANTHER version 16: A revised family classification, tree-based classification tool, enhancer regions and extensive API. Nucleic Acids Research, 49(D1), D394–D403.

Rozenblatt-Rosen, O., Stubbington, M. J., Regev, A., & Teichmann, S. A. (2017). The human cell atlas: From vision to reality. Nature News, 550(7677), 451.

Segerdell, E., Bowes, J. B., Pollet, N., & Vize, P. D. (2008). An ontology for Xenopus anatomy and development. BMC Developmental Biology, 8, 92.

Segerdell, E., Ponferrada, V. G., James-Zorn, C., Burns, K. A., Fortriede, J. D., Dahdul, W. M., Vize, P. D., & Zorn, A. M. (2013). Enhanced XAO: The ontology of Xenopus anatomy and development underpins more accurate annotation of gene expression and queries on Xenbase. Journal of Biomedical Semantics, 4(1), 31.

Van Slyke, C. E., Bradford, Y. M., Westerfield, M., & Haendel, M. A. (2014). The zebrafish anatomy and stage ontologies: Representing the anatomy and development of Danio rerio. Journal of Biomedical Semantics, 5(1), 12.